\documentstyle{mn}
\begin{document}
\def\simlt{\mathrel{\rlap{\lower 3pt\hbox{$\sim$}} \raise
2.0pt\hbox{$<$}}} \def\simgt{\mathrel{\rlap{\lower 3pt\hbox{$\sim$}}
\raise 2.0pt\hbox{$>$}}} \title[Radio Properties of FIRST Radio 
Sources at 1~mJy] {Radio Properties of FIRST Radio Sources at 1~mJy}
\author[M. Magliocchetti, A. Celotti, L. Danese]
{M. Magliocchetti, A. Celotti, L. Danese\\ 
SISSA, Via Beirut 4, 34014, Trieste, Italy} \maketitle
\vspace {7cm }

\begin{abstract} 
This paper presents a detailed analysis of the radio properties for
the sample of faint radio sources introduced in Magliocchetti et
al. (2000). The sample comprises mainly intrinsically low-power
sources whose majority ($\simgt 70$ per cent) is
made of FR~I radio galaxies. These objects show some degree (at 1$\sigma$ 
confidence level) of luminosity evolution, which is also needed to correctly 
reproduce the total number and shape of the counts distribution at 1.4 GHz.
Analysis of the de-evolved local radio luminosity function shows a good 
agreement between data and model predictions for this class of sources. 
Particular care has been devoted to the issue of 'lined' galaxies 
(i.e. objects presenting in their spectra a continuum typical of 
early-type galaxies plus emission lines of different nature), which appear 
as an 
intermediate class of sources between AGN-dominated and starburst galaxies.
Different evolutionary behaviours are seen between the two 
sub-populations of lined and non-lined low-power radio galaxies, the 
first class indicating a tendency for the radio luminosity to decrease with 
look-back time, the second one showing positive evolution. 
We note that different 
evolutionary properties also seem to characterize BL Lacs selected in 
different bands, so that one might envisage an association between lined 
FR~I and the sub-class of BL Lacs selected in the X-ray band.
Lastly, we find evidence for a negligible
contribution of starburst galaxies at these low flux levels. 
\end{abstract}
 

\section{Introduction}
During the last twenty years, attempts to derive models for the
description of the space density and evolution of radio sources have
mainly taken two tacks: the first approach bases its predictions on
the unification paradigm (see e.g. Urry \& Padovani 1995), while the
second one relies on the evolutionary behaviour of the galaxy hosting
the radio source.

Models based on the unification paradigm (Orr \& Brown 1982; Padovani
\& Urry 1992; Maraschi \& Rovetti 1994, Wall \& Jackson 1997 just to
mention a few) stem from the ``relativistic jet'' model by Blandford
\& Rees (1978), where non-thermal continuum radiation is emitted by
plasma relativisticly moving along the jet axis. As a necessary
consequence one has that, as the radio axis gets aligned with the line
of sight, the radio source appears as {\it core-dominated}, i.e.
flat-spectrum beamed radiation dominates the radio emission.  In the
opposite case of misaligned jet axes, the observed flux mainly comes
from the extended isotropically radiating structures (lobes) of the
source, giving rise to steep-spectrum radio emission. It follows that
the latter objects can be considered as the the ``parents'' of
flat-spectrum sources, their different appearance only depending on
the angle between the beaming axis and line of sight.  A further
division can then be made according to the morphology and/or
intrinsic power of the sources: at low radio powers FR~I galaxies (Fanaroff
\& Riley 1974) are assumed to be the parent population of BL Lac
objects, while at higher powers Flat and Steep Spectrum Radio Quasars
(FSRQ/SSRQ) are supposed to be the beamed version of FR~II galaxies.  Note
however that this division is somehow fuzzier (e.g. Antonucci 1993;
Owen \& Ledlow 1994).  According to unification-paradigm based models, 
one therefore has that sources belonging to the same intrinsic
population (e.g. FR~I and BL Lacs) must display the same evolutionary
properties.

A different approach used by a number of authors (Wall, Pearson \&
Longair 1980; Danese et al. 1987; Dunlop \& Peacock 1990; Condon 1984;
Rowan-Robinson et al. 1993) - though still dividing radio sources in
steep and flat spectrum populations - models their evolution and space
density by using descriptions of the epoch-dependent luminosity
function. At variance with the previous case, these models do not
explicitly assume any physical process for the radio-emission and only
focus on the behaviour of the host galaxies.

One of the main limitations affecting these two classes of models (but
especially those deriving from the unification paradigm) is due to the
fact that they were mainly based on datasets including very powerful
sources ($S_{1.4\rm GHz} \simeq 1$~Jy), and their predictions at faint
flux densities greatly diverge because of an inadequate definition of
the low-power tail of the AGN radio luminosity function.  Models
belonging to the second class (Danese et al. 1987; Condon 1984;
Rowan-Robinson et al. 1993) push their analysis down to much lower
flux densities ($S\sim 0.1$~mJy), and assume the contribution to the
radio population at $S\simlt 10$~mJy to be mainly given by a new class
of objects which greatly differ from the radio AGN dominating at
higher fluxes. The nature and level of contribution to the total mJy counts  
of this population is still under
debate. For instance Condon (1984) suggests a population of
strongly-evolving normal spiral galaxies, while others (Windhorst et
al. 1985; Danese et al. 1987) claim the presence of an actively
star-forming galaxy population.

This paper presents an analysis of the radio properties for the sample
of sources introduced in Magliocchetti et al. (2000, hereafter
MA2000). As it will be discussed later on, despite its relative 
smallness, this sample represents the first attempt to derive spectroscopy 
directly from objects uniquely identified as radio-emitting sources (i.e. 
without prior optical identifications). Luminosity functions and luminosity 
distributions will be
derived down to radio-powers $P_{1.4\rm GHz} \sim 10^{20}$ [W
Hz$^{-1}$sr$^{-1}$], therefore allowing a direct comparison with the
models introduced earlier on at such low powers.  We will also analyze
the space density for the different classes of radio-sources and their
relative contribution to the total number counts at flux densities
$1\; {\rm mJy} \simlt S_{1.4\rm GHz}\simlt 10$~mJy.

The layout of this paper is as follows: Section~2 describes the radio,
photometric and spectroscopic properties of the dataset used in our
analysis, while Section~3 deals with the luminosity evolution of radio 
sources and presents the results
for the luminosity function. Section~4 discusses these findings within the 
framework of number counts by comparing observations with model predictions.
Section~5 summarizes
our conclusions.  Throughout this work we will assume $H_0=h_0$ $\cdot$ 100
km s$^{-1}$, with $h_0=0.5$ and $\Omega_0=1$ ($q_0=0.5$).

\section{The Radio Sample}
The sample used for the following analysis has been derived from the FIRST
(Faint Images of the Radio Sky at Twenty centimeters) Survey (Becker,
White \& Helfand 1995) which includes sources down to $\sim $ 0.8~mJy.
The surface density of objects in the catalogue is $\sim 90$ per
square degree, though this is reduced to $\sim 80$ per square degree
if we combine multi-component sources (e.g. sources presenting lobes and 
hot-spots; Magliocchetti et al. 1998). The
catalogue has been estimated to be 95 per cent complete at 2~mJy and
80 per cent complete at 1~mJy (Becker, White \& Helfand 1995).

From this catalogue MA2000 considered 8 regions of approximately 1~deg
each in diameter and performed ``blind'' (i.e. without prior optical
identification of the objects) multi-object spectroscopy by placing
fibers at the positions of 365 radio sources ($\sim 69$ per cent of
the radio sample; this percentage was mainly due to the
geometry of the spectrograph). From these spectra it was possible to
measure 46 redshifts, $\sim 13$ per cent of the targeted objects.  APM
data have then provided photometric information for most of the
sources with measured redshifts. Photometry shows that redshift
measurements were obtained for objects brighter than an apparent
magnitude of R $\simeq 20.5$~mag, and the sample was estimated to be
$\sim$ 100~per~cent complete to R=18.6~mag. This value was derived 
by considering the sources which had photometric measurements but 
lacked of spectroscopic ones as a function of their radio fluxes. It turned 
out that all the objects with R$\le$18.6 were endowed with a redshift 
estimate, independent of the radio flux (Figure 6 in MA2000). 
The above result implies no radio-bias in the acquisition of the 
spectra, the only remaining bias being related to the optical 
properties of the sources.

The objects found with R $\la 20.5$ were a mixture of early-type
galaxies (i.e. absorption systems presenting at most a weak OII
emission line which we will hereafter denote as Early) at relatively high 
redshifts, $z\ga 0.2$ (24, $\sim 52$
per cent of the sample), 'lined' galaxies (absorption systems
presenting strong OII, OIII, H$\alpha$, H$\beta$ emission lines
denoting either ongoing star-formation activity or the presence of an AGN or 
eventually superposition of both these effects) at intermediate redshifts,
$0.02\la z\la 0.2$ (8, $\sim17$ per cent), and very local starburst (SB)
galaxies with $z\la 0.05$ (3, $\sim 6$ per cent). 
By using the diagnostic 
emission line ratios of Rola, Terlevich \& Terlevich (1997), we find 
four of the lined galaxies to show features principally due 
to star-formation (E+SF), while the remaining four to be mainly dominated 
by AGN activity (E+AGN).
Note that our findings are consistent with the existence of a significant 
fraction ($\sim 10$ per cent) of lined objects amongst low power radio
galaxies (Hine \& Longair 1979).  MA2000 also found a number of
broad-lined AGN (type I RL), all at $z\ga 0.8$ (5, $\sim 11$ per
cent of the sample), two narrow-lined AGN (type II RL) (4 per cent),
and 4 stars \footnote{As the objects cannot be classified as
flat/steep radio sources, in the following we name as type I/II Radio
Loud (RL) those objects which are radio--loud (r$>10$, see below) with
broad/narrow emission lines in the optical spectrum.}.  Three objects
(respectively one type II RL, one early-type galaxy and one starburst
galaxy) were subsequently removed from the original sample by
requiring $S_{1.4\rm GHz}\ge$ 1~mJy.  Other seven sources exhibit 
featureless spectra, showing a continuum
emission without the presence of either emission or absorption lines. 
These unclassified objects, endowed with faint (R$\simgt 18.6$) optical 
magnitudes, could be either associated to low-signal-to-noise-spectra 
early-type galaxies or to low-power BL Lacs. As they do not have
redshift determinations, they will only be considered in the last
Section when dealing with the number counts. Their effects on the statistical 
significance of the sample under consideration will instead be discussed in 
the Conclusions.

All the early-type and lined galaxies present absolute magnitudes
distributed about the mean value $M_{\rm R}=-23.1$ with very little
scatter ($\sim 0.5$ at the 2$\sigma$ level) independent of redshift,
implying that passive radio galaxies are reliable standard candles, as
already discussed by e.g. Hine \& Longair (1979). The small spread in
magnitude allows us to determine a very tight R-$z$ relationship for
the objects under exam. From this relation MA2000
estimate $\sim$ 100~per~cent completeness for the spectroscopic sample
with radio fluxes $S_{1.4\rm GHz}\ge$ 1~mJy up to $z=0.3\pm 0.1$. Throughout 
the paper we will therefore adopt $z=0.3$ as the maximum 
redshift for completeness of the sample. This assumption implies that all 
the early-type and lined galaxies with redshifts within $z_{\rm lim}=0.3$ 
will be considered in the subsequent analyses, 
regardless of their optical magnitude. Note that this limit, given 
the spread in the R-$z$ relationship, does not exactly correspond to a 
magnitude cut at R=$18.6$. In fact, we find two sources in the sample 
(both early-type galaxies) which present $z < 0.3$ and R$> 18.6$, even though  
no objects are detected with R$\le 18.6$ and $z > 0.3$. This spread might 
then suggest the existence of $z\le 0.3$ and R$\ge 18.6$ sources, possibly 
corresponding to the unclassified sources we previously referred to, 
which were 
not spectroscopically identified in the MA2000 sample. 
Since the majority of these objects is expected to present redshifts 
$\simgt 0.2$ (sources with R$> 18.6$ and $z\simeq 0.1$ would correspond to 
a $\sim 8\sigma$ event with respect to the R-$z$ relationship and should 
consequently be extremely rare), in order to assess the 
importance of this possible cause of incompleteness, 
the analysis performed in Section 3 will also 
consider two different samples, derived from the original one by respectively 
considering cuts at $z=0.2$ and $z=0.25$. Results obtained in this way will 
be then compared with those derived for $z\le 0.3$ sources.

\begin{figure}
\vspace{8cm}  
\includegraphics{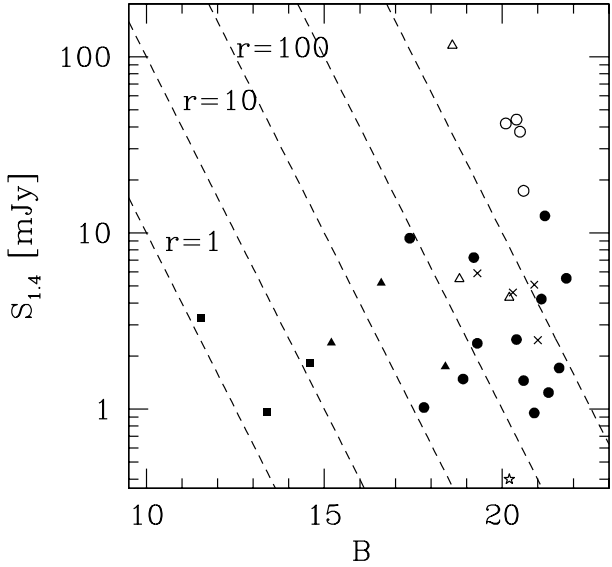}
\caption{Radio flux at 1.4~GHz as a function of B magnitude. Filled
dots identify early-type galaxies, triangles 'lined' galaxies
(filled for E+SF sources, empty for E+AGN objects), filled squares are for 
starbursting galaxies, 
empty circles for type I RL, while stars are for type II RL and crosses for 
unclassified objects (possibly identified as BL Lacs).  
The dashed lines correspond to constant values of the
radio-to-optical ratio $r=1,10,100,1000,10^4$ (see text for details).
\label{fig:F_B}}
\end{figure}

Figure \ref{fig:F_B} shows the distribution of radio fluxes for all
the objects in the sample with measured B magnitudes as a function of
this latter quantity.  The dashed lines indicate different values of
the radio-to-optical ratio $r$, defined as $r=S_{1.4 \rm GHz}\times
10^{\frac{(B-12.5)}{2.5}}$, where S is the radio flux (in mJy) and B
is the apparent magnitude in the blue band.  Note that, according to
the definition of radio-loudness (see e.g.  Urry \& Padovani 1995),
the three starburst galaxies with $r\simlt 10$ cannot in principle be
considered as radio-loud.  Also note that, amongst lined galaxies, 
sources with spectra principally showing signatures due to star formation 
(represented by filled triangles in Figure 1) and sources where 
emission lines mainly originate from AGN activity (empty triangles in Figure 
1) occupy different regions in the B-S plane, the former objects presenting 
radio-to-optical ratios closer to those obtained in the case of starburst 
galaxies. With some confidence we can then conclude that the radio signal 
emitted by this first group of sources mainly stems from intense 
star-formation activity, in close resemblance with the case for starburst 
galaxies.

\begin{figure}
\vspace{8cm} 
\includegraphics{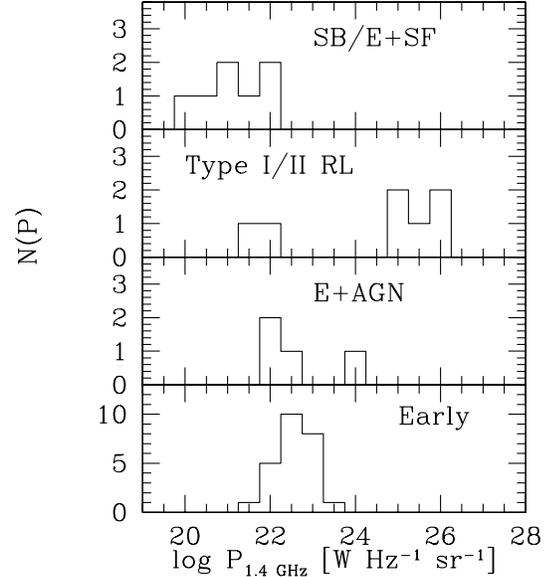}
\caption{Distribution of monochromatic radio power at 1.4 GHz for the
6 classes of sources discussed in the paper.
\label{fig:P_type}}
\end{figure}

Figure \ref{fig:P_type} shows the distribution of monochromatic
radio-power $P_{1.4\rm GHz}=S_{1.4\rm GHz} D^2 (1+z)^{3+\alpha}$ (in
units W Hz$^{-1}$sr$^{-1}$) for the different classes of objects
introduced earlier on. In the above formula $D$ is the angular
diameter distance and $\alpha$ is the spectral index of the radio
emission ($S(\nu)\propto \nu^{-\alpha}$). As we did not have measured
values for this latter quantity, we assumed $\alpha=0.5$ for type I
RL, $\alpha=0.75$ for early-type galaxies (with or
without emission lines due to AGN activity), $\alpha=0.7$ for type II RL and 
$\alpha=0.35$ both for starbursts and E+SF (see e.g.
Oort et al. 1987). In the central panel of Figure \ref{fig:P_type} we
grouped together type I and II RL, type I showing higher radio-power; 
starbursts and E+SF galaxies are also grouped together in the top panel of 
Figure~\ref{fig:P_type}, E+SF sources all appearing for $P_{1.4\rm GHz}\simgt
10^{21}$ [W Hz$^{-1}$sr$^{-1}$].
As one can see from the Figure, early-type galaxies with no emission
lines follow with good approximation the same distribution in power as
those presenting in their spectra emission lines due to AGN activity. Given 
that their radio-powers are always (apart from one
case) $P_{1.4\rm GHz}\simlt 10^{23.5}$ [W Hz$^{-1}$sr$^{-1}$], we
classify these two sub-populations as FR~I galaxies, with the possible 
exception of the source with $P_{1.4\rm GHz}\simeq 10^{24}$ 
[W Hz$^{-1}$sr$^{-1}$]. Note that, as already 
briefly discussed in the Introduction, the distinction between FR~I and FR~II 
galaxies seems to be more complicated and not just determined by the values 
of their radio 
power. As Ledlow \& Owen (1996) have shown, the FR~I/FR~II division is also 
a strong function of the optical luminosity of the host galaxy, optically 
brighter FR~I sources appearing for higher radio powers. Our conclusions for 
these galaxies to all belong to the same class of FR~I is 
however not affected by the previous statement, as the objects in our sample 
present $P < 10^{24}$ [W Hz$^{-1}$sr$^{-1}$], threshold below which only 
members of the FR~I population are found, independent of their magnitude.
From a morphological point of view we find that all the radio images for 
these objects show point-like 
structures. Since the angular resolution of the FIRST survey is $\sim 5$ 
arcsec, corresponding to a physical scale of $\sim 7$~kpc at $z=0.05$ and 
$\sim 35$~kpc at $z=0.3$ for the cosmology adopted in the Paper, one can 
exclude the presence of very extended structures such as those typical of FR 
II sources. This furtherly supports our conclusion for these objects to belong 
to the class of FR~I galaxies.

As expected, the three starburst galaxies show low radio-powers 
($P_{1.4\rm GHz}\simlt 10^{21}$ [W Hz$^{-1}$sr$^{-1}$], while type I RL 
all appear for $P_{1.4\rm GHz}\simgt 10^{25}$ [W Hz$^{-1}$sr$^{-1}$]. 

\section{Statistical and Evolutionary Properties of the Populations}
In principle, in order to directly estimate the evolution of the
different radio populations with time, one has to calculate their
luminosity functions (LF) in different redshift
intervals. Unfortunately though, the sample under consideration does
not include enough sources, and therefore we have to apply the
Schmidt's $V/V_{\rm max}$ test, where $V$ is the comoving volume
`enclosed' by an object and $V_{\rm max}$ is the maximum volume within
which such an object could have been detected above the radio flux and
magnitude limit of the survey. In absence of evolution one has that
the quantity $V/V_{\rm max}$ is uniformly distributed between 0 and 1,
with 0.5 as mean value (Schmidt 1968).

As already stated in Section 2, for the estimate of the quantity 
$\left<V/V_{\rm max} \right>$ we assumed a completeness limit of 
$z_{\rm lim}= 0.3$ in the case of FR~I galaxies (i.e. Early and E+AGN sources) 
with $S\ge 1$~mJy. Errors
for $\left<V/V_{\rm max} \right>$ are then calculated as
$\Delta\left(\left<V/V_{\rm max}\right>\right)=\left({12
N}\right)^{-1/2}$ (Schmidt 1968), where $N$ is the number of objects
belonging to this class. Note that we will not include in our analysis 
either type I/II RL or starburst galaxies (even though we show their 
evolutionary trends in Figure 3 which illustrates the monochromatic power at 
1.4 GHz as a function of redshift), as 
their small numbers do not allow any meaningful statistics; furthermore, 
starburst galaxies do not in principle belong to the population of 
radio-loud sources (see Section 2).

\begin{figure}
\vspace{8cm}  
\includegraphics{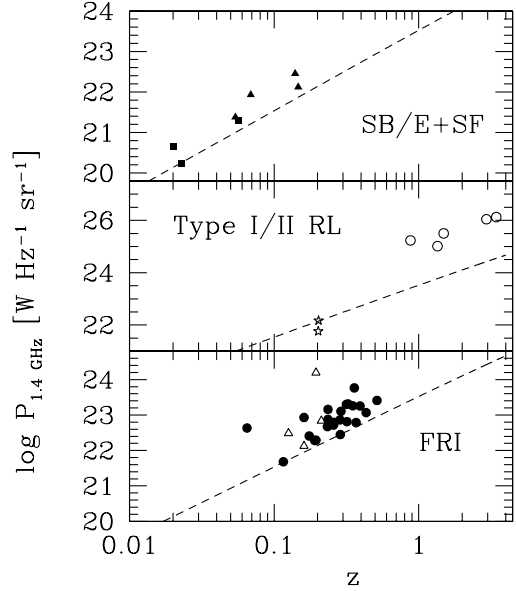}
\caption{Monochromatic power at 1.4 GHz as a function of redshift for
the classes of objects discussed in the paper. In the lower panel,
filled dots are for early-type galaxies, while empty triangles
indicate lined galaxies of AGN origin (E+AGN). In the upper panel, squares 
are for starbursts and triangles for lined galaxies of star-forming origin 
(e.g. E+SF, see text for details). In the middle panel, open circles are for 
type I RL and stars for type II RL. 
The dashed lines show the 1~mJy limit of our sample.
\label{fig:P_z}}
\end{figure}

For the population of FR~I galaxies (18 sources with $z\le 0.3$) 
one gets $\left<V/V_{\rm max}\right>= 0.57\pm 0.06$. Note that this result - 
even though compatible with the hypothesis of no evolution just above the 
1$\sigma$ level - seems to disagree with previous findings (see e.g. 
Urry, Padovani \& Stickel, 1991; Padovani \& Urry, 1992; Maraschi \& Rovetti, 
1994; - where the samples were
extracted from surveys with much higher flux limits, e.g. the 2~Jy
sample (Wall \& Peacock 1985) or the 3CR sample (Laing et al. 1984) - and  
Rowan-Robinson et al., 1993; - where the
sample included objects down to $S_{1.4 \rm GHz}\simeq 0.1$~mJy (Benn et
al. 1993)) for the class of FR~I not to evolve with look-back time.

As already anticipated in Section 2, in order to investigate the stability 
of our results on FR~I galaxies with respect to the eventual presence of 
incompleteness effects in the $z\le 0.3$ sample, we repeated 
the $V/V_{\rm max}$ analysis for different redshift cuts. Even though 
going to lower redshifts 
automatically reduces the number of sources within the sample and the 
statistical significance of the measurements, we nevertheless find  
$\left<V/V_{\rm max}\right>=0.62\pm 0.07$ and 
$\left<V/V_{\rm max}\right>=0.67\pm 0.10$ for objects respectively within 
$z=0.25$ and $z=0.2$, in good agreement with our previous findings, showing 
that incompleteness is not a problem in this case. We can therefore 
confidently conclude that the sample under exam is well suited to examine 
the evolutionary properties of FR~I galaxies. It is also interesting to 
note that the above numbers seem to 
furtherly strengthen the results for positive luminosity evolution of the FR~I
population.    

\begin{table*}
\begin{center}
\caption{Redshift, radio-to-optical ratio and radio power for the lined 
sources described in the text, ordered for increasing power. 
The Table also divides sources into two sub-classes according to their 
spectral features (i.e. whether mainly indicating signatures of 
star-formation activity (E+SF) or the presence of an AGN (E+AGN)).}
\begin{tabular}{lllll}      
Object& Type& Redshift& $r$ & P$_{1.4 \rm GHz}$[W Hz$^{-1}$ sr$^{-1}$] \\
\hline
 2244\_086  & E+SF&  0.054& 28.61  &$2.4\cdot 10^{21}$\\
 2244\_006  &E+SF& 0.069 & 226.6 &$8.58\cdot 10^{21}$\\
 2244\_044  &E+SF& 0.147 & 398.6  &$1.32\cdot 10^{22}$\\
 2236\_012   & E+AGN &0.161&? &$1.33\cdot 10^{22}$\\
 ELAIS2\_017 &E+SF& 0.14 & ?  &$2.82\cdot 10^{22}$\\
 2244\_029  &E+AGN& 0.126 & 1815  &$3.55\cdot 10^{22}$\\
 2236\_023   &E+AGN&0.212 &  5158  &$6.85\cdot 10^{22}$\\
  2236\_013  & E+AGN& 0.195 &3.195e+04 &$1.56\cdot 10^{24}$
\end{tabular}
\end{center}
\end{table*} 

At this point it is worth spending a few lines on the effects of the class 
of lined galaxies on the analysis performed so far. The choice of 
making a distinction between galaxies which presented in their spectra lines 
of mainly stellar origin (E+SF) and those which indicated the presence of an 
AGN in their core (E+AGN) has been based on the relative strength of emission 
line ratios (see Section 2). 
However, it is possible to note that there is indeed a smooth transition 
between these two classes of objects, not only in the relative strength of 
the different emission lines, i.e. whether more due to star formation rather 
than AGN activity, but also in the radio-power, 
radio-to-optical ratio and redshift distribution (see Figures~1,2,3 and 
Table~1) of these 
sources. Also, if one confines the analysis to radio-power and 
radio-morphology only, all these objects would be identified as FR~I galaxies.
In principle one cannot therefore exclude all the lined 
galaxies to be indeed members of the FR~I population, relative emission line 
strengths simply reflecting a passage between more AGN dominated to more 
star-formation dominated sources within the class of FR~I. Note that, in 
general, it is quite common to find ``composite'' galaxies containing both a 
starburst and an AGN (see e.g. Hill et al., 2001).  
In this case, lined FR~I galaxies as a whole would be seen as a ``bridge'' 
(also meant in the sense of decreasing look-back times) 
between AGN-fuelled objects (i.e. those presenting spectra typical of 
early-type galaxies) and sources where the radio signal is dominated by 
processes connected with the intense star-formation activity (starbursts).

The issue connected with these ``transition'' objects is quite delicate as 
their eventual presence or absence within the class of FR~I galaxies can 
modify the apparent evolutionary properties of this population.
In fact, if we consider as FR~I all the lined galaxies (i.e. E+AGN and E+SF) 
and not only those 
where emission lines seem to mainly stem from AGN activity, the value for 
$\left<V/V_{\rm max}\right>$ decreases to $0.52 \pm 0.06$, in this case 
perfectly compatible with no evolution. It follows that 
one has to devote particular care in treating these sources as they might 
influence eventual conclusions. We stress that this effect should not depend 
on the statistical significance of the sample, since lined galaxies are in 
general expected to make up for $\sim 20$ per cent of low-to-intermediate 
redshift radio samples (see e.g. Sadler et al., 1999). 

More stable results are instead those separately related to the two 
sub-populations of Early and E+AGN. For the first class of objects 
in fact we find $\left<V/V_{\rm max}\right>=0.60\pm 0.07$ - compatible with 
{\it positive} evolution at the 2$\sigma$ level -, while 
$\left<V/V_{\rm max}\right>=0.35\pm 0.14$ - suggesting {\it negative} 
evolution - in the case of E+AGN FR~I. 
Even though the paucity of sources here does not allow any strong statement, 
it is nevertheless worth noticing that this negative trend also seems to 
hold if one considers all the lined galaxies together, whatever the 
origin of the lines is, for which $\left<V/V_{\rm max}\right>=0.33\pm 0.10$.  
In general we 
find early-type galaxies to show positive luminosity evolution, while lined 
galaxies tend to evolve in a negative way. The bottom line in this case seems 
to be that, even though belonging to the 
same population of FR~I, these two classes of E and E+AGN sources show quite 
different evolutionary behaviours; their combination gives rise to a 
mildly evolving trend for FR~I galaxies as a whole, but this result might 
hide an internal dichotomy and requires further investigation. 

Note that these different (and opposite) evolutionary trends could mirror 
the behaviours
established for the beamed population of BL Lac objects, respectively
as lower power 'X-ray selected' type and higher power 'radio selected'
sources. In this respect, models which postulate the evolutionary
connection between positively evolving radio--loud quasars into
negatively evolving BL Lacs (Cavaliere \& Malquori 1999) should refer
to the sub-population associated with E+AGN galaxies only.

While keeping the above warnings in mind, in the 
following analysis we will rely on the division made amongst lined galaxies 
introduced in Section 2 when describing the spectroscopic sample. We will then 
consider as belonging to the FR~I population only early-type galaxies 
plus those lined galaxies where emission lines were mainly of AGN origin 
(E+AGN). For this sample we have then estimated the degree of evolution and 
obtained the local radio luminosity function (LF) to be compared with model 
predictions.\\
We parameterize the evolution as $P(z)$$=P(0)\;\rm
exp^{t(z)/\tau}$ (pure luminosity evolution PLE), where $t(z)$ is the
look-back time in units of the Hubble time (defined as 
$t(z)=\int_0^z1/[(1+z)^2(1+\Omega_0\;z)^{1/2}]$) and $\tau$ is the
time-scale of the evolution in the same units; the best values for
$\tau$ are found by requiring $\left<V/V_{\rm
max}\right>=0.5\pm\left(\frac{1}{12 N}\right)^{1/2}$, while the local
power $P_{1.4 \rm GHz}(0)$ can be obtained by inverting the expression
for luminosity evolution. 

\begin{figure}
\vspace{8cm}  
\includegraphics{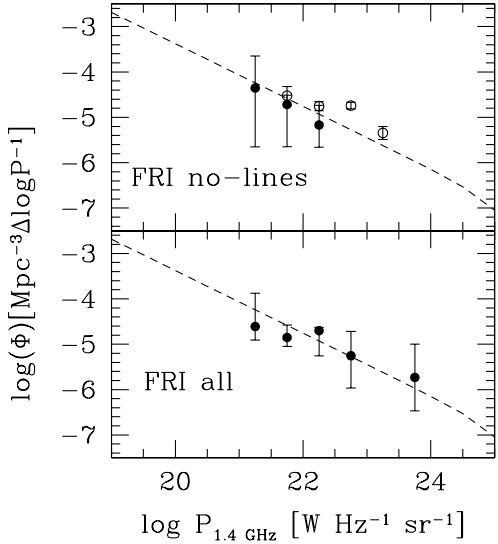}
\caption{Local radio luminosity function at 1.4 GHz for the whole sample of 
FR~I galaxies (lower panel) and the sub-class of FR~I with no emission lines 
in their spectra (Early - upper panel). Error-bars are given by the
sum in quadrature of the Poisson errors and the variations of the
number density associated to a $1\sigma$ change of the evolution
parameter $\tau$. Note that we only include objects with $z\le 0.3$, our
completeness limit, and $S\ge 1$~mJy. Dashed lines indicate the Dunlop \& 
Peacock (1990) model for pure luminosity evolution (see text for details) .  
Filled circles in the
upper panel correspond to the local LF under the hypothesis of
luminosity evolution with $\tau=0.06$, while open circles are obtained
in the case of no evolution (see text for details).
\label{fig:L_z}}
\end{figure}

According to this approach, one then finds $\tau=0.17^{+0.50}_{-0.10}$ for 
FR~I galaxies, compatible with positive evolution at the 1$\sigma$ level.
Note that, by restricting this analysis to the sub-classes of 
early-type and E+AGN galaxies separately, one would get 
$\tau\sim 0.06$ for the first case and $\tau\sim -0.05$ in the second one. 
The statistical significance 
of these results is however limited by the small dimensions of both samples 
together with the limited redshift range ($0\le z\le 0.3$) allowed to 
consider the effects of any eventual evolution. 

The above values for the evolution parameter $\tau$ have then been used to 
derive the local LF by de-evolving at $z$=0 the luminosity of each source
and by subsequently grouping the sources in bins of $\Delta {\rm
log} P=0.5$. The LF has been obtained  according to the expression
\begin{eqnarray}
\Phi(P)=\sum_i N_i(P, P+\Delta P)/V_{\rm max}^i(P),
\end{eqnarray}
where $N_i$ is the number of objects with luminosities between $P$ and
$P+\Delta P$, and $V_{\rm max}^i(P)$ is their maximum volume,
calculated as in Section 3. The values for the LF have then been
corrected by means of the factor $0.8\times 0.66$ to take into account
the 80 per cent completeness of the FIRST survey at 1~mJy (see Becker et al.,
1995) and the percentage of sources with fibres placed on.\\
Figure \ref{fig:L_z} shows the resulting LF (lower panel). 
Following the analysis performed in this Section, FR~I galaxies have
been furtherly divided into lined (E+AGN) and early-type 
galaxies in order to show any systematic difference in the 
LF trend. The result for early-type galaxies is represented by the 
filled circles in the upper panel of Figure \ref{fig:L_z} while, due to the 
paucity of sources, we do not show the LF behaviour in the case of lined FR~I.

The dashed lines indicate the predictions of one of the models of
Dunlop \& Peacock (1990; hereafter DP90), which provides an
alternative way of parameterizing  pure luminosity evolution and assumes
positive luminosity evolution for all steep-spectrum sources, regardless
of their power.  The particular model among those proposed by DP90
(but also see Rowan-Robinson et al. 1993) considers a luminosity
function of the form
\begin{eqnarray}
\Phi(P,z)=\Phi_{\rm SP}(P)+ \Phi_{\rm ELL}(P,z),
\label{eqn:1}
\end{eqnarray}
where
\begin{eqnarray}
\Phi_{\rm ELL}(P,z)=10^{-6.91}/\left\{\left[P_{1.4}/P_{\rm c}(z)\right]^{0.69}+
\left[P_{1.4}/P_{\rm c}(z)\right]^{2.17}\right\}
\label{eq:phi}
\end{eqnarray}
taken at $z=0$ gives the local LF for steep spectrum FR~I+FR~II sources, and 
${\rm log}P_{\rm c}(z)=26.22+1.26 z-0.26 z^2$ (given in W$^{-1}$ Hz$^{-1}$) 
is the evolving ``break'' luminosity.
$\Phi_{\rm SP}(P)$ is the non-evolving LF for
the spiral/irregular galaxy population.  Since all the sources in the FR~I 
sample appear to have spectra with a continuum typical of early-type galaxies, 
we neglect this last term in equation (\ref{eqn:1}) and only consider the
predictions for the population of ellipticals
(i.e. $\Phi(P,z)\equiv\Phi_{\rm ELL}(P,z)$). \\
As Figure \ref{fig:L_z} (filled circles) shows, the agreement between 
data and model is good both in the 
case of the whole sample and for early-type galaxies only (where we have 
multiplied the measured LF by the factor 1/0.78 to account for the percentage 
of early-type sources amongst FR~I galaxies). 

In order to test the effects of luminosity evolution on the
calculations of the local LF, the upper panel of Figure \ref{fig:L_z}
also presents (in open circles) the results obtained for the 
sample of early-type FR~I galaxies under the assumption of negligible evolution
of the population. In this case the data points shift
to the right-hand side of the plot, resulting in an overprediction of
the number of low-power radio sources found locally (e.g. Toffolatti
et al., 1987) and illustrated by the dashed line in
Figure~\ref{fig:L_z}.  

We also note that integration of the DP90 LF in the interval 
$10^{23}\le P\le 10^{25}$ predicts 4$\pm 2$ objects 
to be found in the observed area and for that luminosity range up 
to $z=0.3$, consistent with our findings for very few detections 
of higher-power (FR~II) sources at low-redshifts.  

\section{Number counts}
\begin{figure}
\vspace{8cm}  
\includegraphics{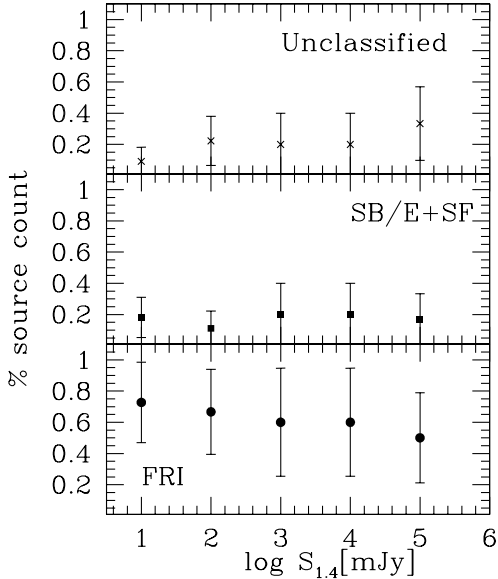}
\caption{Percentage of source counts as a function of flux at 1.4~GHz
for the different classes of objects discussed in the paper. In the middle 
panel, starburst galaxies only appear for $S\simlt 3$~mJy.
\label{fig:density}}
\end{figure}

As a final step we analyze the relative contribution of different
populations to the number counts. Figure \ref{fig:density} shows
the percentage of source counts for the different classes of objects
included in our spectroscopic sample up to $S_{1.4\rm GHz}=6$~mJy (note that 
here we have not included any redshift cut), where the
sample looses statistical meaning due to the paucity of sources.  It
is interesting to note that, for {\it any} value of the flux, the sum of FR~I 
galaxies and unclassified objects accounts for $\simgt 90$ per cent of 
the radio sources with an optical identification,
while starburst galaxies only constitute a small portion of the sample
($\simlt 10$ per cent for $S\simlt 3$~mJy and $z\simlt 0.05$, none beyond 
these values). 
Note that -- as the
procedure for spectra acquisition of the radio sources of this sample
was performed without prior photometric identification -- in principle
MA2000 could have obtained spectroscopic information for sources with
strong emission lines (unless heavily obscured) regardless their
optical brightness. This implies that the sample is more likely to be
biased against FR~I galaxies (in general ellipticals with no or weak
emission lines) than starbursts, which makes our conclusion on the
relative absence of starburst galaxies in the sample even stronger. 
Similar results on the small 
contribution of starburst galaxies to the radio population at mJy level were
obtained by Gruppioni et al. (1998) and Georgakakis et al. (1999).

At this point it is also compelling to examine the constraints on the
evolutionary degree of the FR~I population from the source number
counts. Under the assumption of a luminosity evolution of the form 
${\rm exp}^{t(z)/\tau}$ (see Section 3), integration of the local LF - 
expressed in eq.(\ref{eq:phi}) with $P_c(z)\equiv P_c(0)$ - 
over the luminosities 
$10^{21}\le P\le 10^{24}$ [W Hz$^{-1}$sr$^{-1}$] typical of the FR~I 
population, leads to a total number of
sources between $0\le z\le 3$, on an area of $8\times \pi \theta^2$
(with $\theta=1^\circ$ - see MA2000), respectively of 62, 488 and 7205
for $\tau=10$ (i.e. no evolution), 0.17 and 0.06. These predictions
have to be compared with the value of 656 which is the actual number
of objects from the FIRST survey (corrected for the 80 per cent level
of completeness of the radio sample at 1~mJy) that have been found in
the considered area. Both the values of $\tau$ describing no- and
strong-evolution seem to be ruled out as they respectively grossly 
underpredict and overpredict the observed integrated number
counts, while a better agreement with the data is obtained for $\tau=0.17$.\\ 
Note that the two values which have been ruled out by these observational 
constraints on the integral number counts were those respectively obtained 
by analyzing the evolutionary properties of the sub-class of early-type 
galaxies ($\tau=0.06$) and of the ``enlarged'' population of FR~I ($\tau=10$) 
- where in this case sources were considered to belong to the latter class 
merely on the basis of their radio power and morphology (see Section 3).\\ 
On the other hand, $\tau=0.17$ well describes the shape of the counts 
distribution at 1.4 GHz below $\sim$ 1 Jy (Silva et al., 2001) and - 
together with a LF of the form described by eq.(3) - can reproduce the 
number of FR~I sources (with $P\ge 10^{21}$ [W Hz$^{-1}$sr$^{-1}$] 
and $z\le 0.3$) per redshift interval ($\Delta z=0.1$, see 
Table~2), found for the sample described in this Paper. We therefore
conclude that the same degree of luminosity evolution derived from a 
$\left<V/V_{\rm max}\right>$ analysis of the FR~I population is also 
needed to provide both the correct total number of sources and
the shape of the counts for $S_{\rm 1.4 GHz}\ge 1$~mJy.

\begin{table}
\begin{center}
\caption{Number of FR~I sources (with $P\ge 10^{21}$ [W Hz$^{-1}$sr$^{-1}$]) 
per redshift interval predicted by a LF of the form expressed by equation (3) 
for an evolution parameter $\tau=0.17$ 
as compared with the observational findings from the spectroscopic sample.}
\begin{tabular}{lll}      
Redshift Interval& Predicted& Observed\\
\hline
$0<z\le 0.1$&   1.26& 1 \\
$0.1<z\le 0.2$& 6.3& 7 \\
$0.2<z\le 0.3$& 13.2& 10\\
\end{tabular}
\end{center}
\end{table} 

\begin{figure}
\vspace{8cm} 
\includegraphics{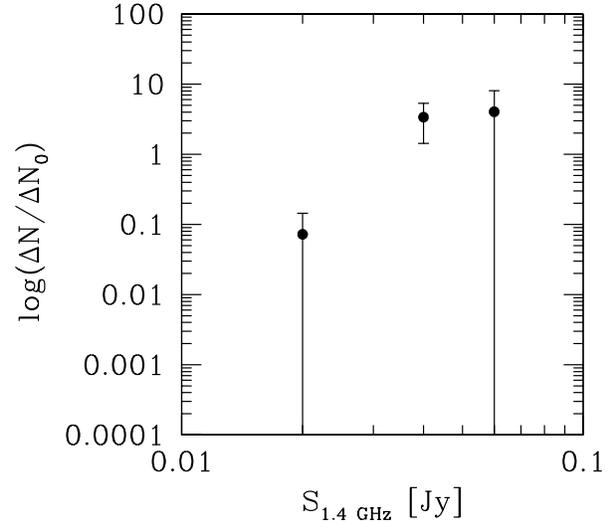}
\caption{The differential number counts for type I RL,
normalized according to Padovani \& Urry (1992) ($\Delta N_0=17(S/{\rm
Jy})^{-2.5}$ Jy$^{-1}$ sr$^{-1}$).
\label{fig:qso}}
\end{figure}

Finally, let us consider the above results in the light of the
findings for beamed objects, according to radio-loud unification
models, as the present sample extends the source counts to
significantly lower fluxes with respect to the $\sim$ Jy samples used
for these analysis so far. In Figure \ref{fig:qso} we report the
differential number counts for type I RL sources. Our findings are in
marginal agreement ($\sim$ 2.5 $\sigma$) with the extrapolation to
lower fluxes of beaming models (e.g. Padovani \& Urry
1992). Furthermore a cut-off in the counts is observed at flux levels
$<$ 40 mJy consistently with the above model predictions.

\section{Conclusions}

In this paper we used the set of data described in MA2000 to derive
the radio-properties of different classes of radio objects at mJy
level.  Our main conclusions are as follows:\\
\\
\noindent
1) The majority of the sample (between $\sim 70$ and $\sim 90$ per cent, 
where the lower limit is 
obtained under the assumption of all the unclassified objects to belong 
to the population of low-power BL Lacs, while the upper limit corresponds to 
the case for all the unclassified objects to be low signal-to-noise-spectra 
early-type galaxies), independent of the
flux level up to $S_{1.4 \rm GHz}\simeq 10$~mJy, is made of FR~I
galaxies. This class includes sources with optical spectra typical of 
early-type galaxies, both with and without emission lines arising from AGN 
activity.\\
2) Within the completeness limit of the sample, the 
population of FR~I as a whole shows some degree of luminosity evolution 
(at 1$\sigma$ 
confidence level). This is also found if one lowers the redshift limit 
for completeness and ensures the stability of the result. The degree of 
evolution is described by an evolution parameter $\tau\sim 0.17$ and can 
correctly reproduce both the integral and differential source counts for 
fluxes $S\ge 1$~mJy.  \\ 
3) Within the population of FR~I, it is found that the sub-class of FR~I 
sources with optical spectra typical of 
early-type galaxies is positively evolving at the 2$\sigma$ confidence level, 
with $\left<V/V_{\rm max}\right>\simeq 0.60$. 
An opposite behaviour is instead seen for lined (E+AGN) FR~I galaxies, for 
which $\left<V/V_{\rm max}\right>\simeq 0.35$. \\
4) Models for the local LF can account for the data both in the case of the whole 
sample of FR~I and for early-type galaxies down to powers 
$P_{1.4\rm GHz}\sim 10^{21}$ [W Hz$^{-1}$sr$^{-1}$].\\
5) Results on type I RL appear to be marginally consistent with both the
number counts and the low flux cutoff predicted by the extrapolation
of Jy level results by beaming models.\\
6) The contribution of starburst galaxies at such 
flux limits is still negligible, of the order of less than $\sim 10$ per cent 
for $S\simlt 3$~mJy and $z\simlt 0.05$, decreasing to zero beyond these values.
This is to be expected if one considers the distribution of radio-to-optical 
ratios for the different mJy to sub-mJy surveys available in literature. A 
comparative analysis performed by Prandoni et al. (2001), in fact shows the 
population of star-burst galaxies to dramatically increase for decreasing 
$S\le 1$~mJy radio fluxes {\it and} very bright optical magnitudes 
($I\le 17.5$).\\
\\

Within this scenario, the population of lined galaxies plays a 
very important 
role. A sharp distinction between galaxies with spectra dominated by 
signatures of AGN activity and those showing features due to star formation 
has been proved not to be an easy task (see e.g. Rola, Terlevich \& 
Terlevich, 1997). This is mainly caused by the existence of numerous 
``composite'' galaxies containing both an event of intense star-formation 
and an AGN (Hill et al., 2001). In the sample presented in this work, there
seems to be a smooth transition (in terms of radio power, radio-to-optical 
ratio and even redshift) between lined galaxies of AGN and star-forming 
origin, 
so that this class of objects as a whole could be seen as a bridge between 
AGN-dominated sources and pure starbursts.\\  
One might then envisage a connection with the
apparent cosmological trend amongst galaxies. In particular the low
activity galactic nuclei, possibly associated with lower efficiency,
might be associated with the presence of ongoing star formation giving
rise to late-type host galaxies at lower redshifts. On the contrary,
higher nuclear powers would plausibly quench the processing of gas
into stars at earlier epochs.  

Note that different evolutionary properties seem
to characterize BL Lacs selected in different bands: X--ray selected
lower power objects (whose synchrotron emission peaks at higher
frequencies; Giommi \& Padovani 1994; Fossati et al. 1998) which are
negatively evolving, and more powerful radio selected BL Lac (peaking
at lower frequencies) consistent with no or marginally positive
evolution (e.g. Urry \& Padovani 1995). 

The above conclusions (especially those in 2 and 3) obviously strongly rely 
on the completeness and/or statistical significance of the sample under 
consideration. As already discussed in Section 2, a possible cause of 
incompleteness might stem from the existence of R$>18.6$ and $z\le 0.3$ 
sources -- possibly associated with some of the seven unclassified objects in 
MA2000 -- for which no spectroscopic identification was possible.\\
Due to the procedure for spectra acquisition performed by MA2000, we can 
reasonably assume these objects to belong to the class of early-type galaxies, 
as the presence of emission lines of whatever origin would have allowed 
redshift measurements for objects much fainter than R=18.6. These early-type 
galaxies would then most likely exhibit redshifts $\simgt 0.2$, since they 
are not expected to differ by more than a factor $\sim 6\sigma$ from the 
R-$z$ relationship found for passive radio galaxies.
 This implies all these sources to show values $V/V_{\rm max}>0.5$. It 
therefore follows that the eventual presence of these objects in our sample 
would have just strengthened the conclusions for a positive evolution of 
early-type galaxies, while leaving the findings for a negative evolution of 
the lined (E+AGN) FRI population unaltered. Furthermore, 
results obtained for the (mild) 
evolution of the FR~I population are confirmed by a completely independent 
analysis performed on the effects of evolution on the integral and 
differential number counts for sources with $S\ge 1$~mJy. And also, 
our findings 
on the radio luminosity function are not only in agreement with model 
predictions, but also with results from much wider samples (see e.g. 
Magliocchetti et al., 2001).

While the above 
discussion assesses the goodness of the sample described in this work, it is 
nevertheless clear that it only constitutes a small fraction of the 
observable radio sources. This stresses the need for more and better data 
-- probing intermediate-to-high redshifts -- 
to come from future surveys in order to achieve more precise and quantitative 
conclusions on extremely important issues for both radio astronomy and 
cosmology such as the evolution of the FR~I population and its connection with 
different BL Lacs flavour.

\vskip 0.5 truecm
\noindent 
{\bf Acknowledgments} S.J. Maddox is thanked for extremely interesting
discussions.  The Italian MURST is acknowledged for financial support.

\end{document}